\documentclass[prd,twocolumn,showpacs,preprintnumbers,amsmath,amssymb]{revtex4}

\usepackage{graphicx}
\usepackage{dcolumn}
\usepackage{bm}
\usepackage{longtable}
\usepackage{afterpage}

\newcommand{\pipi}{$\pi\pi$ }
\newcommand{\mpipi}{$m_{\pi\pi}$ }
\newcommand{\osdrdf}{$OSDR_{DF}$ }
\newcommand{\osdrdfn}{$OSDR_{DF}$}
\newcommand{\mps}{\mpi^2}
\newcommand{\mpi}{m_\pi}
\newcommand{\re}{\mbox{Re }}
\newcommand{\im}{\mbox{Im }}
\newcommand{\ra}{\rightarrow}
\newcommand{\PPint}{-\hspace{-0.502cm}\int}
\newcommand{\sps}{\,\,\,\,\,}

\begin{document}

\title{Dispersion relations with crossing symmetry for \mbox{\boldmath $\pi\pi$} \mbox{\boldmath $D$} and \mbox{\boldmath $F$} wave amplitudes}

\author{R.~Kami\'nski}
\affiliation{
Department of Theoretical Physics
Henryk Niewodnicza\'nski Institute of Nuclear Physics,
Polish Academy of Sciences,
31-342, 
Krak\'ow, Poland}

\begin{abstract}
A set of once subtracted dispersion relations with imposed crossing symmetry condition 
for the \pipi $D$- and $F$-wave amplitudes is derived and analyzed. 
An example of numerical calculations 
in the effective two pion mass range from the threshold to 1.1
GeV is presented.
It is shown that these new dispersion relations impose quite strong  constraints on the analyzed \pipi interactions
and 
are very useful tools to test the \pipi amplitudes.
One of the goals of this work is to provide a complete set of
equations required for easy use. 
Full analytical expressions 
are presented.
Along with the well known dispersion relations successful in testing the \pipi $S$- and $P$-wave amplitudes, those presented here for the $D$ and $F$ waves give a complete set of tools for analyzes of the \pipi interactions.

\end{abstract}

\maketitle

\section{Introduction}

The introduction of crossing symmetry conditions into dispersion relations
for the $\pi\pi$ amplitudes
was promoted by Roy in 1971 \cite{Roy1971}.
Since then, now well-known the Roy's equations have been successfully used to test the $\pi\pi$ $S$- and $P$-wave amplitudes \cite{oldRoy,Pennington73,Caprini:2005zr,Descontes,Kaminski:2002,Bern,A4,CCGL,KPYIII,KPYIV,KPYTallahassee}.
These equations with two subtractions have been used for example to analyze the low energy \pipi
interaction parameters \cite{Descontes} and to eliminate the long standing "up-down" ambiguity in scalar-isoscalar $\pi\pi$ wave amplitudes \cite{Pennington73,Kaminski:2002}.
Quite recently interest in these equations has increased significantly due to series of works by Bern and Madrid groups \cite{Bern,A4,CCGL,KPYIII}.
In these analyzes, authors have used, inter alia, the Roy's equations to construct
the \pipi  $S$ and $P$ wave amplitudes fulfilling crossing symmetry for
effective two pion mass \mpipi from the threshold to over 1~GeV. 
To perform it they had to describe the \pipi $S$-, $P$-, $D$-, $F$- and $G$-wave amplitudes using phenomenological parameterizations below 1.42 GeV \cite{KPYIII} or 2 GeV \cite{A4} and Regge amplitudes up to several GeV.
In such a way these analyzes delivered prescription for constructions of unitary \pipi amplitudes for many partial waves
in very wide energy ranges.
Together with the outcome of new precise data near the \pipi threshold \cite{ThrData} these works
led also to very accurate determination of the mass and width of the $f_0(600)$ (or $\sigma$)
resonance and of the threshold parameters \cite{Caprini:2005zr,KPYTallahassee}.

Very recently the \pipi $S$ and $P$ wave amplitudes with crossing symmetry constraints have been analyzed using not only the Roy's equations but also once subtracted dispersion relations \cite{KPYTallahassee,KPYIV}.
Due to one less subtraction 
these new equations, called GKPY, proved to be much more
demanding than the Roy ones.
Above around \mpipi = 450 MeV the GKPY equations have much smaller uncertainties
so they impose stronger constraints on the \pipi amplitudes.

In the works mentioned above only the $S$ and $P$ wave
amplitudes have been directly
fitted to the Roy's or GKPY equations. 
The higher partial wave amplitudes have been fitted only indirectly via their relations
with the $S$ and $P$ waves within the dispersive equations.

In this paper are derived
once subtracted dispersion relations with imposed crossing symmetry condition 
for 
$D$ and $F$ waves, hereafter called \osdrdfn.
Together with the Roy's and GKPY equations they will form a complementary set of dispersion relations which can be very useful in testing the \pipi $S$-, $P$-, $D$- and $F$-wave amplitudes.

This paper is organized as follows: in Section II general structure and analytical properties of 
the \osdrdf
are presented and discussed.
Individual components of these equations are
analyzed in detail.
The third section contains an example of numerical calculations.
Full derivation of \osdrdf and analytical expressions for their components
are given in Appendices A and B.
Discussion of results and summary are in Section IV.

Through the text, partial waves with orbital momentum $l$ and isospin $I$ are denoted by $WI$ or just by $W$ if the isospin does not need to be specified.
The $W$ can be $S$, $P$, $D$, $F$ or $G$ for $l$ equal to 0, 1, 2, 3 or 4, respectively.

\section{Dispersion relations}
\label{Section:Theory}

Dispersion relations, for example the Roy's and GKPY ones, 
relate real parts of given partial wave amplitudes with sets of 
imaginary parts of other ones.
General form of dispersion relations with one subtraction for the $D$- and $F$-wave amplitudes
reads:
\begin{equation}
    \begin{array}{lll}
    \hspace{-0.cm}\mbox{Re } f_{\ell}^{I}(s)  = 
    \displaystyle 
     -\frac{1}{24}(a^0_0 - \frac{5}{2} a^2_0) \delta_{I1} \delta_{l3}+ \\
    \hspace{-0.cm} 
        \displaystyle \sum\limits_{I'=0}^{2}
        \displaystyle \sum\limits_{\ell'=0}^{3}
     \,\,\,\, \PPint\limits \limits_{4m_{\pi}^2}^{s'_{max}}\hspace{-0.1cm} ds'
     {  K_{\ell \ell^\prime}^{I I^\prime}(s,s')} {\mbox{Im }f_{\ell'}^{I^\prime}
     (s')} + {  d_{\ell}^{I}(s)}
\end{array}
\label{DF_OSDR}
\end{equation}
where $s = m_{\pi\pi}^2$.
The first component $-(a^0_0 - 2.5 a^2_0)/24$ is so called subtracting term $ST_\ell^I$.
Two remaining parts are called the kernel $KT_\ell^I(s)$ and driving $DT_\ell^I(s)$ terms.

Below $s' = s'_{max}$ partial wave  amplitudes $f^{I}_l(s)$ can be expressed by their relations with the experimental \pipi phase shifts $\delta_l^{I}(s)$ and inelasticities $\eta_l^{I}(s)$ 
  \begin{equation}
f^{I}_l(s)=\frac{\eta_l^{I}(s)e^{i\delta_l^{I}(s)}-1}{2i\sigma(s)}
  \label{Eq:ampl} 
  \end{equation}  
where
  \begin{equation}
\sigma(s) = \sqrt{\frac{s-4m_{\pi}^2}{s}}.
  \label{Eq:sigma} 
  \end{equation}
For brevity hereafter the real parts of amplitudes on the left side of 
Eq. (\ref{DF_OSDR})
will be called "output amplitudes" and the imaginary parts of amplitudes 
on the right side - "input amplitudes".

The angular momentum $l'$ of the input amplitudes goes from 0 to 3 (the $S$, $P$, $D$ and $F$ waves).
As was shown in \cite{KPYIV} the input amplitude for the $G$ wave ($l^\prime=4$) is very small and has negligible influence on the output.
Because of the Bose symmetry 
the sums $l^\prime+I^\prime$ and $l + I$ 
for input and output amplitudes respectively, must be even.

As presented in Appendix A the subtracting terms $ST_\ell^I$ in once subtracted dispersion relations are constant and are determined by 
combinations of real parts of partial wave amplitudes at the $\pi\pi$ threshold.
As can be seen from threshold expansion 
\begin{equation}
Re f_l^I(k) = k^{2l}\Big(a_l^{I}+b_l^{I} k^2+O(k^4)\Big)
\label{Eq:thrExp}
\end{equation}
where the pion momentum $k = \sqrt{(s/4-m_{\pi}^2)}$,
the only nonzero amplitudes at $k = 0$ are those for the $S0$ and $S2$ waves.
Their values at the threshold are scattering lengths $a_0^0$ and $a_0^2$, respectively.
In Appendix A it is shown that 
in the case of dispersion relations 
for the $D$ and $F$ partial waves the only nonzero combination of these two scattering lengths in the subtracting terms is for the $F$ wave.
It is due to nonzero integral (from 0 to 1) of the Legendre polynomial for this partial wave. 
The scattering lengths $a_0^0$ and $a_0^2$
can be treated as an input and
may be fixed using for example ChPT predictions (see e.g. \cite{A4}) or can be fitted to data and to theoretical constraints (see e.g. \cite{KPYIII,KPYTallahassee,KPYIV}).

In the kernel terms $KT_\ell^I(s)$ the products of the input amplitudes and of the kernels $K_{ll'}^{II'}(s,s')$, defined in Appendix A, are integrated over $s^\prime$ 
from the \pipi threshold to $s'_{max}$ i.e. up 
to the energy where the phenomenological parameterizations of the phase shifts and inelasticities in input amplitudes (see Eq. (\ref{Eq:ampl})) are quite well known.
In practice $s'_{max} \sim 1.4-2$~GeV (see e.g. analyzes of the Roy's \cite{A4} and GKPY equations \cite{KPYIV}).  

Above $s^\prime = s'_{max}$ the input amplitudes are parameterized using 
the Regge formalism.
Integrals in this $s'$ range are grouped into the driving terms $DT_\ell^I(s)$.
Contrary to the kernel terms, the products of the input amplitudes and corresponding kernels in $DT_\ell^I(s)$ must be doubly integrated - over $s'$ and $t$.
This is due to the $t$ dependence of the Regge amplitudes.

As shown in Appendix A, the only singularities in Eq. (\ref{DF_OSDR}) are those at $s^\prime = s$ in the diagonal kernel elements i.e for $l = l^\prime$.
Therefore one has to take principal value there.

As was already pointed out in \cite{KPYIV} while comparing the GKPY and Roy's equations for the $S$ and $P$ waves, an essential feature of the once subtracted dispersion relations is their slower convergence than in twice subtracted dispersion relations.
In the case of one subtraction, the integrands in the $KT_\ell^I(s)$ and in $DT_\ell^I(s)$ behave as $1/s'^2$ for $s' \to \infty$ while in
the Roy's equations as $1/s'^3$.
Due to this difference the input amplitudes at higher energies in the former equations enter with higher weights than in the latter ones.
It is especially important for higher partial waves which contribute mainly above 1 GeV and therefore are undervalued in relations with two subtractions.
Since the output amplitudes for given $l$ partial wave depend also on the input ones with $l'=l$ this argument is reinforced in this analysis dealing with $D$ and $F$ output partial wave amplitudes. 

Constant value of the $ST_\ell^I$ in \osdrdf leads to constant value of its errors which,
one can expect, are much smaller (for higher $s$) than these which would be in analogous relations with two subtractions.
It is due to the fact that subtracting terms in these relations are not constant but are linear functions of $s$ which leads to increase of their uncertainties with energy.
Detailed comparison of the errors in the once and twice subtracted dispersion relations for the $S$ and $P$ waves (the GKPY and Roy's equations, respectively) can be found in \cite{KPYIV}.

The application range of the \osdrdf, Eq. (\ref{DF_OSDR}), is the same as of the GKPY equations in \cite{KPYIV} and comes  about 1.1 GeV.
One can expect that below this energy only the $D0$ output amplitude will show a clear increase with the energy due to presence of the $f_2(1270)$ resonance. 
The absence of any resonance in the $D2$ wave and the relatively large mass of meson $\rho_3(1690)$ in the $F1$ one, lead to rather small variation of amplitudes in these waves below 1.1 GeV.

\begin{table}[h!]
\hspace{.0cm}\begin{tabular}[b]{c|c|c|c|c}  
Wave  & TE & $ST_\ell^I$ & $KT_\ell^I(s) $ & $DT_\ell^I(s)$  \\ 
\hline
$D0$ & 0 & $0$ & $\alpha_0 + c_0\beta_0$ & $\gamma_0 + d_0\delta_0$ \\
$F1$ & 0 & $-\frac{1}{24}(a_0^0-\frac{5}{2}a_0^2)$ & $A + \alpha_1 + \beta_1$ & $B + \gamma_1 + \delta_1$  \\
$D2$ & 0 & $0$ & $\alpha_2 + c_2\beta_2$ & $\gamma_2 + d_2\delta_2$ \\
\end{tabular}
\caption{Comparison of the threshold expansion TE (Eq. (\ref{Eq:thrExp}))  results for $s \to 4m_{\pi}^2$
and of the threshold behavior of subtracting $ST_\ell^I$, kernel $KT_\ell^I(s)$ and driving $DT_\ell^I(s)$ terms for $D$ and $F$ partial waves. 
The $\alpha_I$ and $\gamma_I$ are terms of the order $O(s-4m_{\pi}^2)$ and $\beta_I$, $\delta_I$ of the order 
$O(s-4m_{\pi}^2)^2$. Values of the $A$, $B$, $c_I$ and $d_I$ constants are explained in the text.
}
\label{Table:thr}
\end{table}
In practical applications of the \osdrdf it is very interesting and useful 
to compare the threshold behavior of the $ST_\ell^I$, $KT_\ell^I(s)$ and $DT_\ell^I(s)$ for $D$ and $F$ waves.
According to the threshold expansion (\ref{Eq:thrExp}), 
the sum of these components 
should vanish at $s = 4m_{\pi}^2$ for all these waves.
Comparison of the threshold expansions of the subtracting, kernel and driving terms is presented in Table \ref{Table:thr}.
For the $D0$ and $D2$ waves the zero order parts in all components are equal to zero, which automatically 
ensures correct behavior of the full output amplitudes at $s = 4m_{\pi}^2$.
Of course sum of the first order parts $\alpha_I$ and $\gamma_I$ must be equal to zero and sum of the
$c_I$ and $d_I$ should give corresponding scattering lengths. 
In case of the $F1$ wave the nonzero value of subtracting term must be canceled
by sum of nonzero values of $KT_\ell^I(4m_{\pi}^2)$ and $DT_\ell^I(4m_{\pi}^2)$.
Sums of the first and second order parts should give zero.

It is worthy to emphasize here that such cancellations demand strong and proper mutual relations between 
the amplitudes for all waves integrated over very wide energy range in the $KT_\ell^I(s)$ and $DT_\ell^I(s)$. 
These relations for the zero order parts of the $F1$ wave, i.e. for $A$ and $B$ in Table \ref{Table:thr},
involve also the threshold parameters of the lowest partial waves ($S0$ and $S2$) and
are expressed by the well known Olsson sum rule \cite{Olsson}
 \begin{widetext}
  \begin{eqnarray}
    \label{Eq:LowEnergy} 
    \nonumber     (a_0^0-\frac{5}{2}a_0^2)&=&2\int\limits_{4m_{\pi}^2}^{s'_{max}}\frac{ds'}{\pi s'(s'-4m_{\pi}^2)}
    \Bigg[ 2 Im f_0^0(s') +10 Im f_2^0(s') 
    +\hspace{-0.1cm}9 Im f_1^1(s') + 21 Im f_3^1(s') - 5 Im f^2_0(s') - 25 Im f_2^2(s') \Bigg] \\
     &+&  12\int\limits_{s'_{max}}^{\infty}\frac{Im F_t^1(s',0)}{\pi s'(s'-4m_{\pi}^2)} ds' \\ \nonumber
&&\text{\hspace{-2.25cm}where $F_t^1(s',t)$ is the isospin 1 amplitude in the $t$ channel (see Appendix A).}
  \end{eqnarray}
 \end{widetext}
Of course an identical relation may also be obtained from the  GKPY equations.
In practice 
this sum rule gives a chance to verify an accuracy of used 
parameterizations of many input partial wave amplitudes in a very wide energy range
and validates the choice (or result of a fit) of scattering lengths
$a_0^0$ and $a_0^2$.


\section{Numerical analysis}
\label{Section:Results}
As an example of practical application of the \osdrdfn, the $D$ and $F$ output amplitudes have been calculated using  the input ones from \cite{KPYIV}. 
These input amplitudes were obtained there in dispersive data analysis using forward dispersion relations, sum rules, the Roy's and GKPY equations.
The $D$ and $F$ waves have not been directly fitted to any dispersion relation.

\begin{figure}[h!]
  \centering
\includegraphics[scale=0.46]{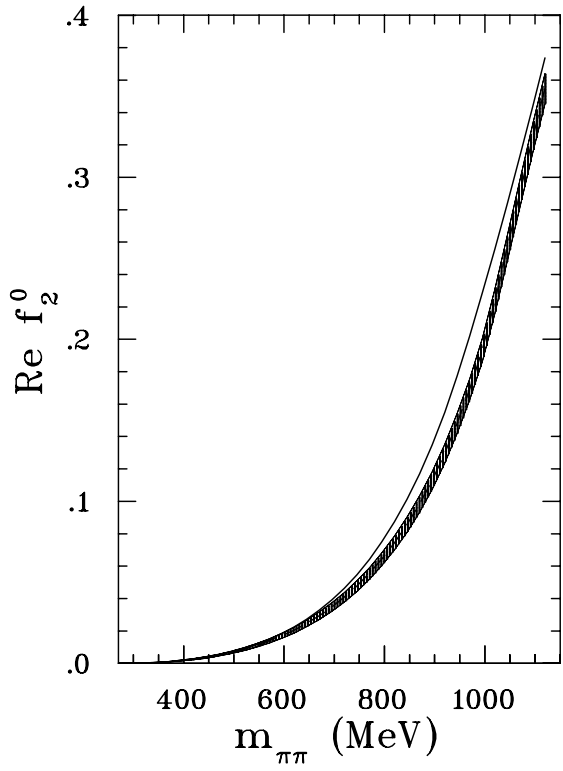}
  \caption{Input (solid line) and output (dashed line) for the $D0$ wave amplitude in Eq. (\ref{DF_OSDR}). 
  Gray band represents errors of the output.}
\label{fig:InOut.D0}
\end{figure}

\begin{figure}[h!]
  \centering
\includegraphics[scale=0.46]{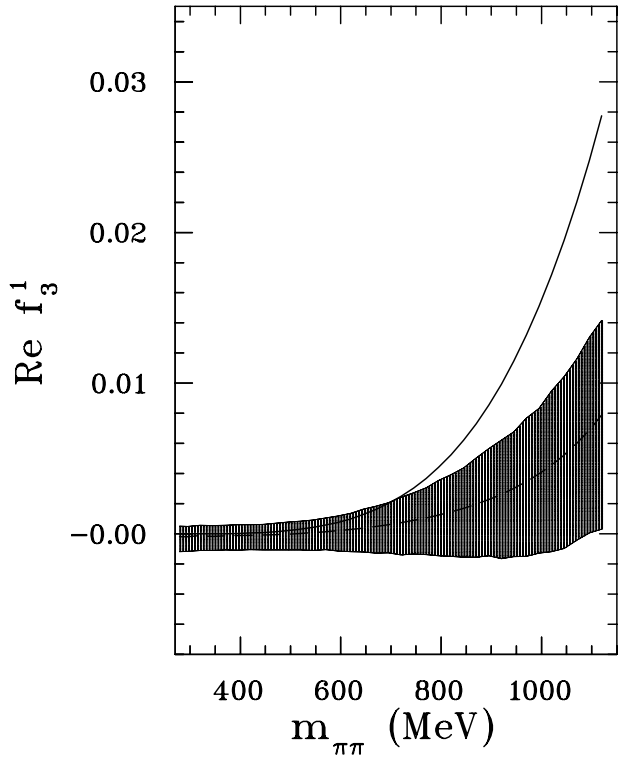}
  \caption{As in Fig. \ref{fig:InOut.D0} but for $F1$ wave amplitude.}
\label{fig:InOut.F1}
\end{figure}

\begin{figure}[h!]
  \centering
\includegraphics[scale=0.46]{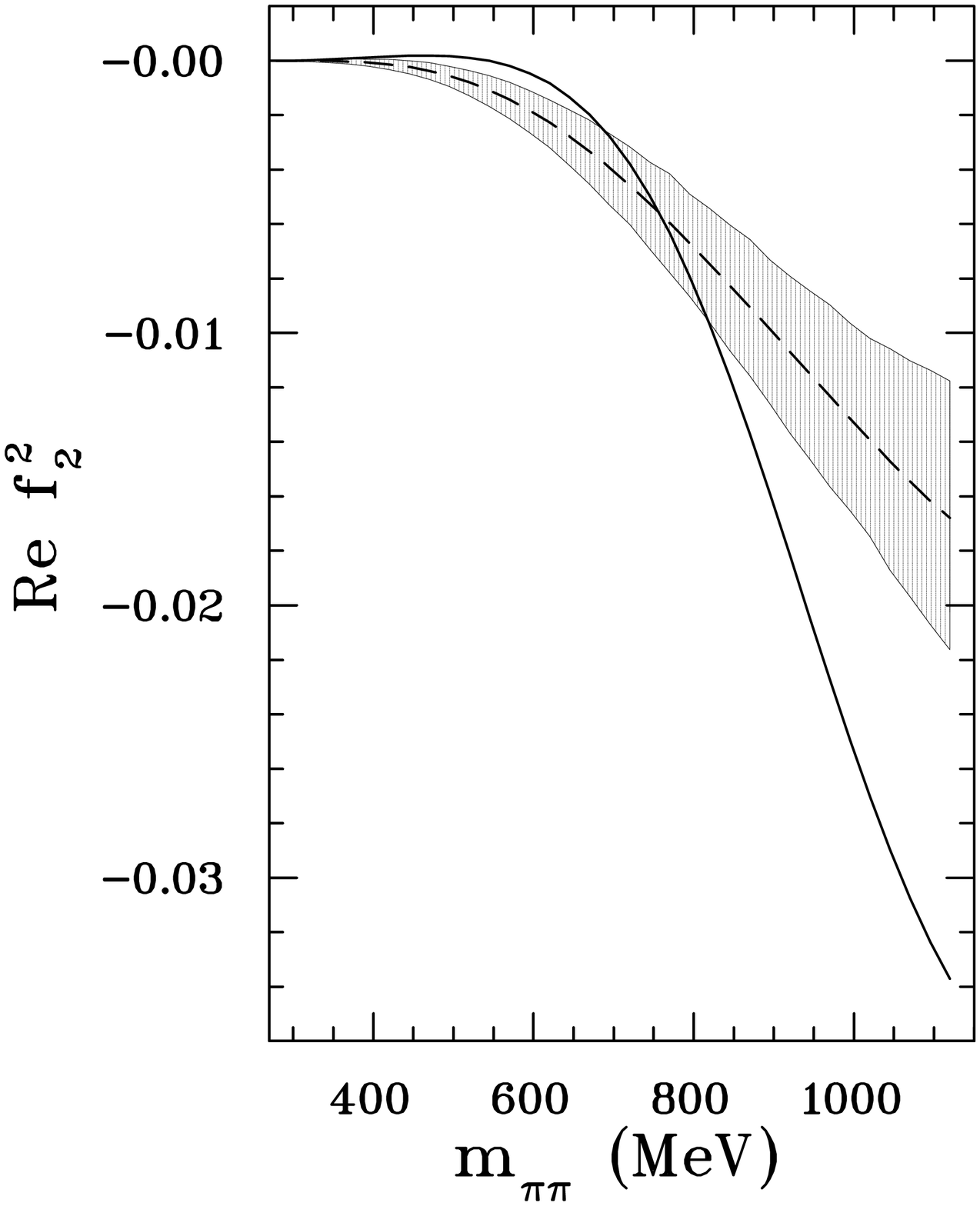}
  \caption{As in Fig. \ref{fig:InOut.D0} but for $D2$ wave amplitude.}
\label{fig:InOut.D2}
\end{figure}

Figures \ref{fig:InOut.D0} to \ref{fig:InOut.D2} present \mpipi distributions of the output amplitudes
and the real parts of the amplitudes whose imaginary parts have been used as the inputs in Eq. (\ref{DF_OSDR}).
The error bands of the output amplitudes have been calculated using Monte Carlo method for all 53 parameters 
used in \cite{KPYIV} to parameterize the low and high energy behavior of 
all 
input amplitudes. 
Assuming Gaussian distributions of these parameters they have been varied randomly $10^3$ times 
within their left and right $3\sigma$ which resulted in three Gaussian distributions for the three output amplitudes.
This procedure has been carried out for 25 values of $s$ between the \pipi threshold and 1110 MeV.
The errors of the output amplitudes have been determined by widths of the Gaussian functions fitted to the left and right sides of these distributions at each $s$ independently.
Of course, due to the large number of parameters, these errors are very similar to those 
obtained as square root of sum of squares of individual errors of each parameter. 
However, the Monte Carlo method has been chosen to present possible asymmetries of final errors caused by correlations between varied parameters.

As is seen on the figures \ref{fig:InOut.D0} to \ref{fig:InOut.D2} the error bands of the output $D0$ and $D2$ wave amplitudes go to zero for $s \to 4m_{\pi}^2$.
The nonzero errors of the $F1$ wave at the threshold are due to the nonzero value of subtracting term and in fact represent its error.
In Table \ref{Table:thr} this constant $ST_\ell^I$ is a
linear combination of two scattering lengths of the $S0$ and $S2$ waves.
Therefore, the full error of the output  $F1$ wave amplitude at the threshold 
is completely determined by the uncertainties of these two threshold parameters. 
Convergence of the $D0$ and $D2$ output amplitudes to 0 for 
$s \to 4m_{\pi}^2$ is well seen.
In the case of the $F1$ wave, the output amplitude goes to 
$(-1 \pm 7)\cdot 10^{-4}$ which is compatible with the predicted zero value in Table~\ref{Table:thr}.
The error has been calculated in the same way as the errors of the output amplitudes.


\begin{figure}[h!]
  \centering
\includegraphics[scale=0.46]{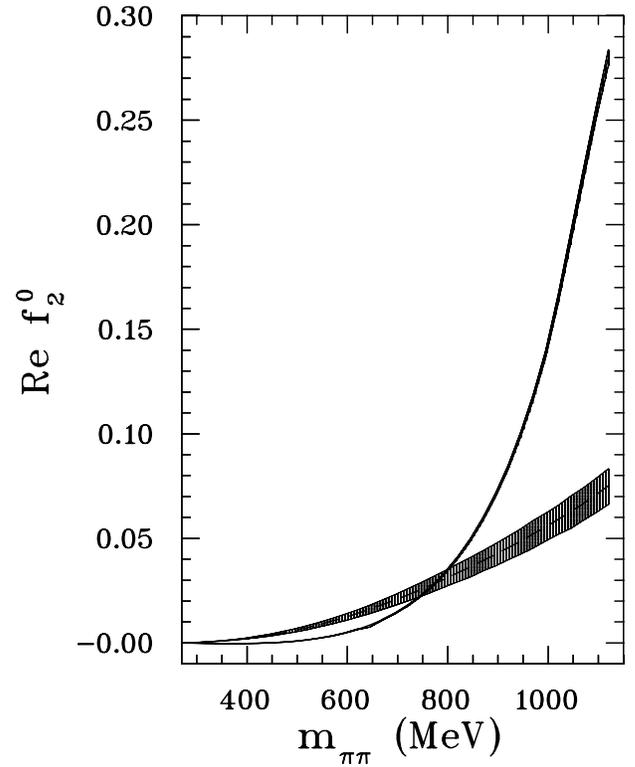}
  \caption{Components of the output $D0$ wave amplitude: kernel term (solid line) and driving term (dashed line). Gray bands represent their errors.}
\label{fig:KTDT.D0}
\end{figure}
\begin{figure}[h!]
  \centering
\includegraphics[scale=0.44]{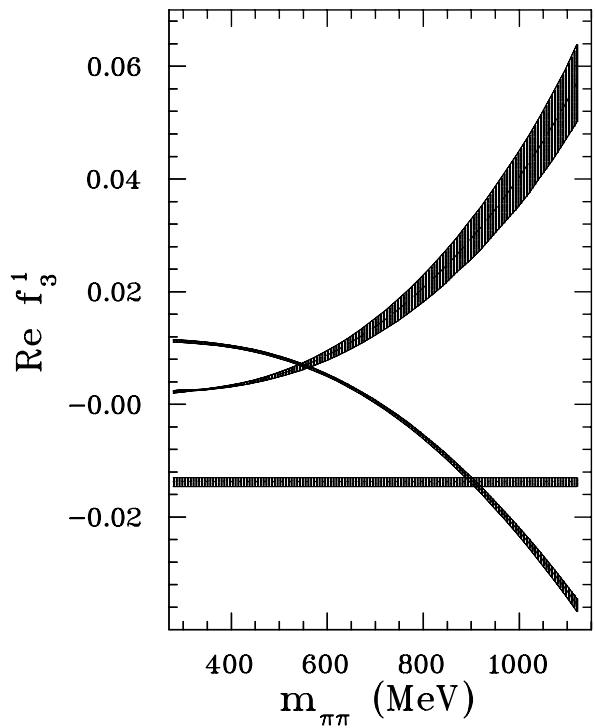}
  \caption{As in Fig. \ref{fig:KTDT.D0} but for the $F1$ amplitude. Horizontal dashed-dotted line represents subtracting term.}
\label{fig:KTDT.F1}
\end{figure}
\begin{figure}[h]
  \centering
\includegraphics[scale=0.44]{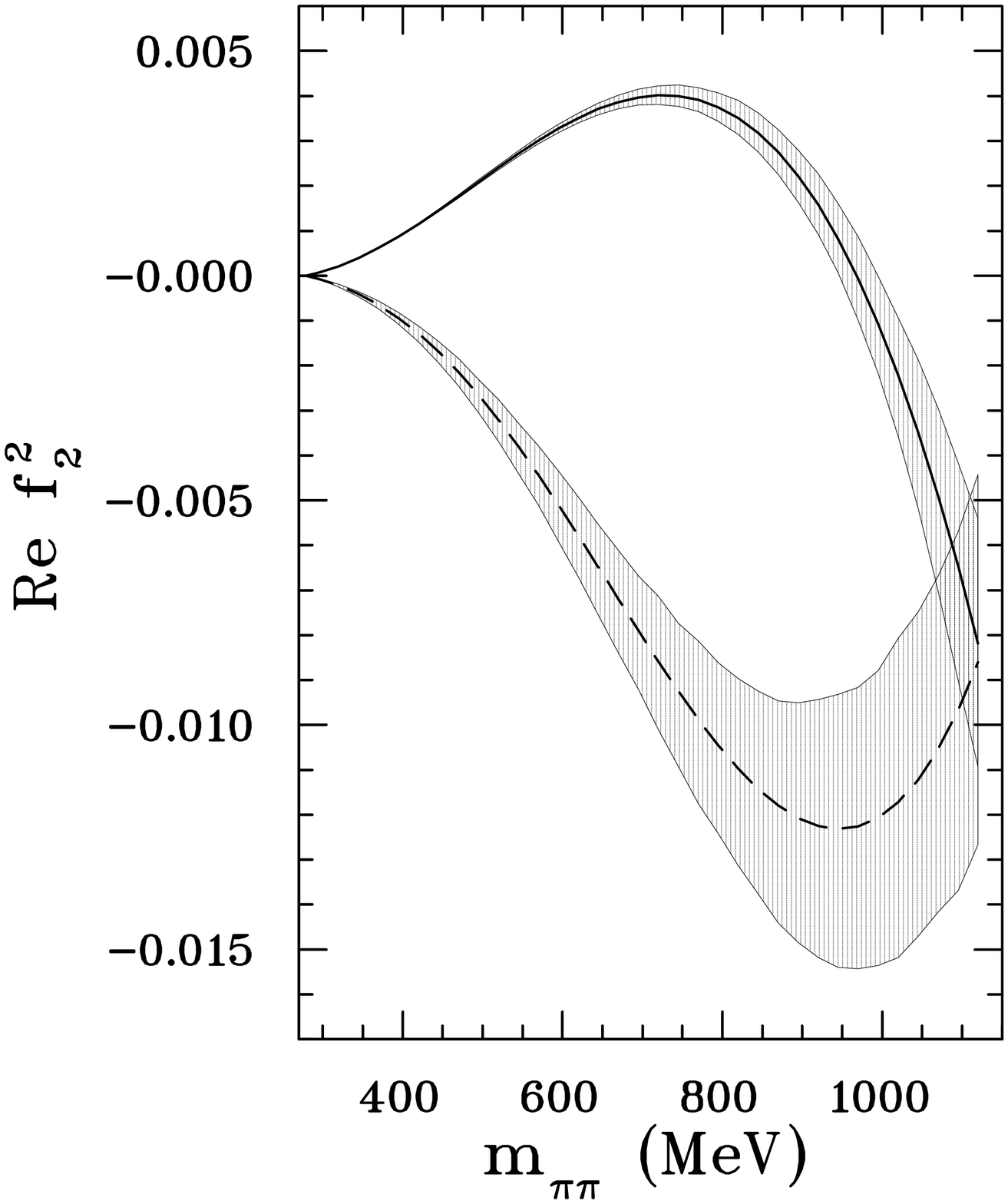}
  \caption{As in Fig. \ref{fig:KTDT.D0} but for the $D2$ wave amplitude.}
\label{fig:KTDT.D2}
\end{figure}

As the $D$ and $F$ wave amplitudes have not been directly fitted to any dispersion 
relation in \cite{KPYIV}, the fact that their input and output amplitudes for $m_{\pi\pi}>800$~MeV differ by more than 
one $\sigma$ is not surprising.
One can expect, however, that 
the use of the presented here dispersion relations in more complete analysis of the \pipi amplitudes e.g. in the analysis of the Bern or Madrid group
would decrease this difference significantly.
It would also increase the weight of the theoretical constraints imposed 
on these and on other wave amplitudes, and therefore would diminish their 
uncertainties.

Vastly different scales in the figures for the $D0$ and for $D2$ and $F1$ waves reflect significant differences
in the sizes of the amplitudes.
As was explained in Section \ref{Section:Theory} these differences
are due to the presence of the meson $f_2(1270)$ in the $D0$ wave close to the studied 
$s$ region and lack of such states in the $D2$ and $F1$ waves.

Figures. \ref{fig:KTDT.D0} to
 \ref{fig:KTDT.D2} present the $m_{\pi\pi}$ dependences of the $ST_\ell^I$, $KT_\ell^I(s)$ and $DT_\ell^I(s)$ components for the $D$ and $F$ wave amplitudes.
Error bands have been calculated for each term separately as for the output amplitudes in Figs. \ref{fig:InOut.D0} to \ref{fig:InOut.D2}.  
Again, due to the presence of the $f_2(1270)$ resonance 
much larger values of the
$KT_\ell^I(s)$ and $DT_\ell^I(s)$ for the $D0$ wave than 
for the $D2$ and $F1$ ones are visible (note different scales on the figures).

For all three waves the errors of the kernel terms are significantly smaller than those of the driving ones.
This is due to the fact that the input amplitudes in $KT_\ell^I(s)$ are more precisely known than those in $DT_\ell^I(s)$.
The errors of the latters are determined by errors of the input amplitudes above $s'_{max}$. 
These amplitudes are, however, less precisely known than those at lower energies.
In particular, small errors of the kernel for the $D0$ wave amplitude 
are due to presence of the well known resonance $f_2(1270)$. 

In Fig. \ref{fig:KTDT.F1} for the $F1$ wave, the straight line represents 
nonzero subtracting term value of $-0.0137^{+0.0007}_{-0.0009}$.
Its errors are completely determined by the uncertainties of the scattering lengths 
of the $S0$ and $S2$ waves.
If the amplitudes of these waves at the threshold are not fixed but fitted 
(as it is e.g. in \cite{KPYIV}) then of course, these uncertainties become functions of the errors of amplitude parameters.
According to what was shown in Table \ref{Table:thr}, the kernel and driving terms also have nonzero values at the threshold to compensate the subtracting term. 
Threshold values of the kernel and driving terms are: $A = 0.0113 \pm 0.0002$ and $B = 0.0023 \pm 0.0002$, respectively.

\section{Conclusions}

A set of once subtracted dispersion relations with imposed crossing symmetry 
condition for the \pipi $D$- and $F$-wave amplitudes has been derived and analyzed.
Analytical structure of these equations and of their components have been studied and described in detail.
It was shown that integrals in these equations converge slower than 
integrals in the twice subtracted dispersion relations (e.g. in the Roy's ones for the $S$ and $P$ partial waves).
Thanks to this, the input amplitudes are less suppressed at higher energies which
is essential for higher partial waves becoming significant only above 1 GeV.
One can expect that use of the one subtraction will lead to the 
generation of smaller uncertainties
of the output amplitudes than those created using two subtractions.
In the latter case, the errors of the subtracting terms, being not constant but linear functions of $s$, grow with $s$.
Therefore, the one subtracted dispersion relations provide more demanding tests for the \pipi amplitudes. 

Apart of the derivation of these equations, a first practical application has been presented. 
The input amplitudes in the low and high energy region have been taken from 
fit to experimental data and from direct, for the $S$ and $P$ wave amplitudes, and indirect fit to dispersive theoretical constraints in \cite{KPYIV}.
It has been shown that because of the $D$- and $F$-wave amplitudes were not fitted directly to any theoretical constraints there, their inputs and outputs, calculated in this paper, differ sizable above about 800 MeV.
One can expect, however, that the compatibility between them will be 
significantly improved if presented here dispersion relations are used in fits of the \pipi amplitudes.

It was shown that nonzero value of the subtracting term in the $F1$ output amplitude creates opportunity to relate low and high energy 
behavior of many partial wave amplitudes with the threshold parameters of the lowest ones.
Therefore one can expect that these equations may help to decrease uncertainties of the $\pi\pi$ amplitudes and errors of these threshold parameters.
One can also expect that the theoretical constraints imposed by these equations 
directly on the $D$ and $F$ wave amplitudes and indirectly on lower ones can lead to slightly more precise determination of the $f_0(600)$ and $f_0(980)$ parameters.

In conclusion the above derived and analyzed set of equations is easy to use 
and can be very helpful in future analyzes of the $\pi\pi$ interactions.
Together with dispersion relations for the $S$ and $P$ waves - the Roy's and GKPY equations with two and one subtraction,
respectively, they can be used as complementary set of equations to 
test $\pi\pi$ amplitudes from the threshold to about 1.1 GeV.

\section{Acknowledgments}

The author thanks B. Loiseau for discussions and for help in editing the
text. 
This work has been partly supported by the Polish Ministry of Science and
Higher Education (grant No N N202 101 368). 


\appendix
\section{Derivation of the once subtracted dispersion relations
}

Let us define following isovector for the scattering amplitudes $F^I(s,t)$ 
of isospin $I$
 \begin{equation}
    \label{eq:F-vector}
    \mbox{\boldmath $\vec{F}$}(s,t) = 
    \left(
    \begin{array}{c}
     F^{0}(s,t) \\
     F^{1}(s,t)  \\
     F^{2}(s,t) 
     \end{array}
     \right).
  \end{equation}
Then a once subtracted dispersion relations can be expressed by
  \begin{eqnarray}
    \label{Eq:DRcuts}
&&\re \mbox{\boldmath $\vec{F}$}(s,t) =  \re \mbox{\boldmath $\vec{F}$}(s_0,t)
    + \frac{s-s_0}{\pi} \\ \nonumber
  &&\times\Bigg[\,\,\,\PPint\limits_{4\mps}^\infty ds'
    \frac{\im \mbox{\boldmath $\vec{F}$}(s',t)}{(s'-s_0)(s'-s)} 
    + -\hspace{-0.45cm}\int\limits_{-t}^{-\infty} ds'
    \frac{\im \mbox{\boldmath $\vec{F}$}(s',t)}{(s'-s_0)(s'-s)}\Bigg]
  \end{eqnarray}
where $s_0$ is the subtraction point to be defined later.

The first and second integral in Eq.~(\ref{Eq:DRcuts}) are taken on the real $s$ axis
along the right and left hand cuts of $F^{I}(s,t)$, respectively.
For any $s$ values along these cuts, one should take the principal values for 
these integrals (see e.g. the discussion in Appendix B of \cite{KPYIV}).

Performing the substitution
  \begin{equation}
  u' = 4m_{\pi}^2 - s' - t
  \end{equation}
and using the crossing symmetry relation
  \begin{equation}
  \mbox{\boldmath $\vec{F}$}(u',t) = \mbox{\boldmath $\hat{C}$}_{su} \mbox{\boldmath $\vec{F}$}(s',t),
  \label{Eq:Cst} 
  \end{equation}
the left hand cut integrals can be recast in terms of the right hand cut ones
  \begin{eqnarray}
    \label{Eq:DR}  
    &&\re \mbox{\boldmath $\vec{F}$}(s,t) = \re \mbox{\boldmath $\vec{F}$}(s_0,t) + \frac{s-s_0}{\pi}  \\ \nonumber
    &&\times\PPint\limits_{4\mps}^\infty ds'
    \Bigg[
      \frac{\im \mbox{\boldmath $\vec{F}$}(s',t)}{(s'-s_0)(s'-s)}
    - \mbox{\boldmath $\hat{C}$}_{su}\frac{\im \mbox{\boldmath $\vec{F}$}(s',t)}{(s'-u_0)(s'-u)}
    \Bigg] \quad 
  \end{eqnarray}
with $u = 4m_{\pi}^2 - s - t$, $u_0 = 4m_{\pi}^2 - s_0 - t$ and with 
the crossing symmetry matrix $\mbox{\boldmath $\hat{C}$}_{su}$
defined in Eq. (\ref{Eq:CSM}).

Using the $s\longleftrightarrow t$ crossing symmetry relation 
one can express the subtracting terms $F^{I}(s_0,t)$ by
  \begin{equation}
  \mbox{\boldmath $\vec{F}$}(s_0,t) = \mbox{\boldmath $\hat{C}$}_{st} \mbox{\boldmath $\vec{F}$}(t,s_0).
  \label{Eq:FCstF} 
  \end{equation}
Following now the same procedure which was used in derivation of the 
$\text{Re} F^{I}(s,t)$ in Eq. (\ref{Eq:DR}), one can get similar, once subtracted dispersion relations for $F^{I}(t,s_0)$: 
  \begin{eqnarray}
    \label{Eq:DRwithCst}  
   &&\re \mbox{\boldmath $\vec{F}$}(t,s_0)=\re \mbox{\boldmath $\vec{F}$}(t_0,s_0)
    + \frac{t-t_0}{\pi}  \\ \nonumber
    &&\times\PPint\limits_{4\mps}^\infty ds'
    \Bigg[
      \frac{\im \mbox{\boldmath $\vec{F}$}(s',s_0)}{(s'-t_0)(s'-t)} 
    - \mbox{\boldmath $\hat{C}$}_{su}\frac{\im \mbox{\boldmath $\vec{F}$}(s',s_0)}{(s'-u_{00})(s'-u_0)}
    \Bigg]  \quad
  \end{eqnarray}
with 
$u_{00} = 4 m_{\pi}^2 - s_0 - t_0$ and $t_0 = 4 m_{\pi}^2 - s_0 - u$. 
The crossing matrices $\mbox{\boldmath $\hat{C}$}_{su}$ and  
$\mbox{\boldmath $\hat{C}$}_{st}$
used in Eqs. (\ref{Eq:Cst}) to (\ref{Eq:DRwithCst}) read
 \begin{equation}
  \label{Eq:CSM}
   \mbox{\boldmath $\hat{C}$}_{su} =
   \begin{pmatrix}
     \frac{1}{3} & -1 & \frac{5}{3} \\
     -\frac{1}{3} & \frac{1}{2} & \frac{5}{6} \\
     \frac{1}{3} & \frac{1}{2} & \frac{1}{6}
   \end{pmatrix},
   \quad
   \mbox{\boldmath $\hat{C}$}_{st} =
   \begin{pmatrix}
     \frac{1}{3} & 1 & \frac{5}{3} \\
     \frac{1}{3} & \frac{1}{2} & -\frac{5}{6} \\
     \frac{1}{3} & -\frac{1}{2} & \frac{1}{6}
   \end{pmatrix}.
 \end{equation}
It is worthy to mention here that the set of equations (\ref{Eq:DRwithCst}) 
can be easily obtained from (\ref{Eq:DR}) using the following replacements:
$s \to t$, $t \to s_0$, $s_0 \to t_0$, $u \to u_0$ and
$u_0 \to u_{00}$.

The problem of the convergence of integrals (\ref{Eq:DR}) and (\ref{Eq:DRwithCst}) has been analyzed and described in \cite{KPYIV}.
It was shown there that, due to the Pomeron contribution from the $I_t=0$ channel, the integrals along the left and right hand cuts are divergent when taken separately.
However, taking both cuts into account simultaneously,
a cancellation occurs and the integrands decay as $1/s'^2$ when
$s'\ra\infty$ which ensures convergence of the integrals.
In the Roy's equations with two subtractions corresponding integrands behave like $1/s'^3$  which ensures convergence of each integral 
separately and of course faster, than in case of once subtracted equations, 
convergence of their sum.

Substituting now (\ref{Eq:FCstF}) and (\ref{Eq:DRwithCst}) into (\ref{Eq:DR}) and following the Roy's original choice: $t_0=4\mps$ and $s_0 = 0$ one gets
\begin{widetext}
  \begin{eqnarray}
    \label{eq:gkpy:final-dr}
    \re \mbox{\boldmath $\vec{F}$}(s,t) & = & \omega\,\mbox{\boldmath $\hat{C}$}_{st}\mbox{\boldmath $\vec{a}$} \\ \nonumber 
     &+& 
    \frac{t-4\mps}{\pi}\PPint\limits_{4\mps}^\infty ds'
    \mbox{\boldmath $\hat{C}$}_{st}
    \left[
      \frac{\im \mbox{\boldmath $\vec{F}$}(s',0)}{(s'-t)(s'-4\mps)} -
      \frac{\mbox{\boldmath $\hat{C}$}_{su}\im \mbox{\boldmath $\vec{F}$}(s',0)}
      {s'(s'-u_0)}\right]
      \\ \nonumber 
       &+& 
    \frac{s}{\pi}\,\,\PPint\limits_{4\mps}^\infty ds'
    \left[
      \frac{\im \mbox{\boldmath $\vec{F}$}(s',t)}{s'(s'-s)}
      - \frac{\mbox{\boldmath $\hat{C}$}_{su} \im \mbox{\boldmath $\vec{F}$}(s',t)}
      {(s'-u_0)(s'-u)}           
    \right]
   \label{Eq:ABCDE}    
  \end{eqnarray}
\end{widetext}
where $\omega$ is some constant (for example $32\pi$ in \cite{A4} and $8/\pi$ in \cite{KPYIV}) and
  \begin{equation}
    \label{eq:a-vector}
    \mbox{\boldmath $\vec{a}$} \stackrel{def}{=} \frac{\mbox{\boldmath $\vec{F}$}(4\mps,0)}{\omega} = 
    \left(
    \begin{array}{c}
     a^0_0 \\
     0 \\
     a^2_0
     \end{array}
     \right).
 \end{equation}
is a vector with, the elements of which are scattering lengths of the $S0$ and $S2$
wave amplitudes and are defined by the threshold expansions given in Eq. (\ref{Eq:thrExp}).

Projection of the 
vector $\mbox{\boldmath $\vec{F}$}(s,t)$ on partial waves $f^{I}_l(s)$ is given by
  \begin{equation}
  \label{Eq:PWP} 
  \mbox{\boldmath $\vec{f}$}_l(s) = \frac{1}{\omega}\int\limits_0^1 d x P_l(x) \mbox{\boldmath $\vec{F}$}(s,t),
  \end{equation}  
where  
 \begin{equation}
    \label{eq:f-vector}
    \mbox{\boldmath $\vec{f_l}$}(s) = 
    \left(
    \begin{array}{c}
     f_l^{0}(s) \\
     f_l^{1}(s)  \\
     f_l^{2}(s) 
     \end{array}
     \right),
  \end{equation}
$P_l(x)$ are Legendre polynomials and 
 \begin{equation}
    \label{eq:x-t}
    t = \frac{(s-4m_{\pi}^2)(x-1)}{2}.
  \end{equation}
Here, due to the symmetry of the integrands, the integration limit was taken from 0 to 1 instead of from~-1~to~1.

In the projection of the isospin amplitudes on partial waves it is convenient to split integrals in Eq. (\ref{eq:gkpy:final-dr}) into low (for $s<s'_{max}$) end high energy part ($s>s'_{max}$).
Value of $s'_{max}$ is determined by limited range of applicability of phenomenological 
parameterizations for the input amplitudes. 
Typically is $s'_{max} \sim 1.4 - 2$~GeV (see for example \cite{A4} and \cite{KPYIV}).

In the low energy part the input amplitudes on the right hand side of (\ref{eq:gkpy:final-dr}) are expressed by 
  \begin{equation}
 \label{Eq:F-f} 
 \mbox{\boldmath $\vec{F}$}(s,t) =  \omega\,\sum_{l}(2l+1) P_l(x) \mbox{\boldmath $\vec{f_l}$}(s) 
  \end{equation}

Analytical expressions for the output partial wave amplitudes are obtained from 
the relations (\ref{eq:gkpy:final-dr}), (\ref{Eq:PWP}) and (\ref{Eq:F-f}) and can be shortly written as 
  \begin{eqnarray}
    \label{eq:DROSDR} 
    \re \mbox{\boldmath $\vec{f}$}_\ell(s) &=& \xi_l \,\mbox{\boldmath $\hat{C}$}_{st} \mbox{\boldmath $\vec{a}$} \\ \nonumber 
    &+& \sum_{l'}\,\PPint\limits_{4\mps}^{s'_{max}} ds' 
    \mbox{\boldmath $\hat{K}$}_{\ell\ell'}(s,s')\im \mbox{\boldmath $\vec{f}$}_{\ell'}(s') + \re \mbox{\boldmath $\vec{f}$}_\ell^{\,\,h.e.}(s)
  \end{eqnarray}
where $\xi_l = \int\limits_{0}^{1} dx P_l(x)$ and  $\re \mbox{\boldmath $\vec{f}$}_\ell^{\,\,h.e.}(s)$ is the hight energy part defined later.
For simplicity all integrals in this equation are grouped in
  \begin{eqnarray}
    \mbox{\boldmath $\hat{K}$}_{\ell\ell'}(s,s') & = & (2\ell'+1) \\
   & \times& \Bigg\{ K_{\ell\ell'}(s,s') \mbox{\boldmath $\hat{1}$}
    - L_{\ell\ell'}(s,s') \mbox{\boldmath $\hat{C}$}_{su}\nonumber\\
   & +&  M_\ell(s,s') \mbox{\boldmath $\hat{C}$}_{st} - N_\ell(s,s')  \mbox{\boldmath $\hat{C}$}_{st}\mbox{\boldmath $\hat{C}$}_{su}\Bigg\}. \nonumber
  \end{eqnarray}  

The $K_{ll'}(s,s'),\,\,\,L_{ll'}(s,s'),\,\,\,M_{l}(s,s')\,\,\, \text{and}\,\,\, N_{l}(s,s')$  kernels read
\begin{eqnarray}
\hspace{-0.5cm}K_{ll'}(s,s')&=&\frac{s}{\pi s'(s-s')}
\int\limits_{0}^{1}\hspace{-.0cm}dx P_l(x)P_{l'}\left(y\right), \label{KernelK}\\
\hspace{-0.5cm}L_{ll'}(s,s'))&=&\frac{s}{\pi}
-\hspace{-0.40cm}\int\limits_{0}^{1}dx P_l(x)
\frac{P_{l'}\left(y\right)}
{u'\left(u'-s\right)}, \label{KernelL} \\
\hspace{-0.5cm}M_l(s,s')&=&\frac{1}{\pi(s'-4m_{\pi}^2)}\int\limits_{0}^{1}dx P_l(x)
\frac{t-4m_{\pi}^2}{s'-t}, \label{KernelM} \\\nonumber
\text{and\,\,\,\,\,\,\,\,\,\,\,\,\,\,\,} && \\
\hspace{-0.5cm}N_l(s,s')&=&\frac{1}{\pi s'}-\hspace{-0.4cm}\int\limits_{0}^{1}dx P_l(x)
\frac{4m_{\pi}^2-t}{u'} \label{KernelN}
\end {eqnarray}
where  $y = (u'-t)/(u'+t)$.
Integrands of the kernels $L_{ll'}(s,s')$ and $N_l(s,s')$ have singularities at $u'=0$ 
which for $s'<\frac{1}{2}(s+4m_{\pi}^2)$ is in the range of integration. Therefore principal values are taken for these integrals.

The full analytical expressions for the $K_{ll'}^{II'}$ elements of 
$\mbox{\boldmath $\hat{K}$}_{\ell\ell'}(s,s')$ are presented in Appendix B.

In the high energy parts of the integrals in Eqs. (\ref{eq:gkpy:final-dr}), 
the input amplitudes can be expressed by $t$-channel ones $\mbox{\boldmath $\vec{F_t}$}(s,t)$
using the Regge parameterizations.
The relation between amplitudes in the $s$ and $t$ channels is
\begin{equation}
 \label{Eq:Amp_s_t} 
 \mbox{\boldmath $\vec{F}$}(s,t) = \mbox{\boldmath $\hat{C}_{st}$} \mbox{\boldmath $\vec{F_t}$}(s,t).
\end{equation}
Following it, one can conclude that high energy partial wave amplitudes derived directly from Eqs. (\ref{eq:gkpy:final-dr}) read
\begin{widetext}
  \begin{eqnarray}
    \label{eq:Output_he}
    \re \mbox{\boldmath $\vec{f}$}_\ell^{\,\,h.e.}(s) &=&
    -\hspace{-0.51cm}\int\limits_{s'_{max}}^{\infty} ds' \Bigg\{
    \frac{s}{\pi}\int\limits_{0}^{1} dx P_l(x) 
    \Bigg[\frac{\mbox{\boldmath $\hat{1}$}}{s'(s'-s)} - \frac{\mbox{\boldmath$\hat{C}_{su}$}}{(s'-u_0)(s'-u)}\Bigg]    
    \mbox{\boldmath $\hat{C}_{st}$} Im \mbox{\boldmath $\vec{F}_t$}(s',t)
    \\ \nonumber
&+& \mbox{\boldmath $\hat{C}_{st}$} 
\Bigg[ \left( M_l(s',s)\mbox{\boldmath $\hat{1}$} - N_l(s',s) \mbox{\boldmath $\hat{C}_{su}$}\right)
\mbox{\boldmath $\hat{C}_{st}$} Im \mbox{\boldmath $\vec{F}_t$}(s',0)  \Bigg]
\Bigg\}.
  \end{eqnarray}
\end{widetext}

Finally, the full expression for given output partial wave amplitude can be written as in Eq. (\ref{DF_OSDR}):
  \begin{equation}
  Re f_\ell^{I}(s) = ST_\ell^I + KT_\ell^I(s) + DT_\ell^I(s) 
  \end{equation}
where for the subtracting term:
  \begin{eqnarray}
  \hspace{-0.9cm}ST_\ell^I &=& \xi_l \,\mbox{\boldmath $\hat{C}$}_{st} \mbox{\boldmath $\vec{a}$} = -\frac{1}{24}(a^0_0 - \frac{5}{2} a^2_0) \delta_{I1} \delta_{l3},\\\nonumber
&&\hspace{-1.9cm}\text{and for the kernel terms:}\\
\label{KTIl}
   \hspace{-0.9cm}KT_\ell^I(s) &=& \sum_{I'=0}^2 \sum_{\ell'=0}^{3}    
   \hspace{0.2cm} \PPint\limits_{4\mps}^{s'_{max}} ds'  {K}_{\ell\ell'}^{I I'}(s,s') \im f_{\ell'}^{I'}(s').
  \end{eqnarray}
The driving terms $DT_\ell^I(s)$ are given by Eq. (\ref{eq:Output_he}). 


\section{Analytical expression for kernels}

Following work \cite{wanders} for the Roy's equations, one can express eighteen kernels ${K}_{\ell\ell'}^{I I'}(s,s')$ in Eq. (\ref{KTIl}) as
functions of the four kernels $K_{ll'}(s,s')$ and eight combinations of the 
$L_{ll'}(s,s')$, $M_l(s,s')$ and  $N_l(s,s')$ ones 
from Eqs. (\ref{KernelK}) to (\ref{KernelN})
\begin{equation}
\begin{array}{lllllll}
K^{00}_{20} &= &K_{20} - I_{20}/3,&\,\,\,\,\,\,\,\,&
K^{02}_{20} &= &-\frac{5}{3}I_{20}, \\ \nonumber
K^{01}_{21} &= &3I_{21},&&
K^{00}_{22} &= &5(K_{22} - \frac{1}{3}I_{22}), \\ \nonumber
K^{02}_{22} &= &-\frac{25}{3}I_{22},&&
K^{01}_{23} &= &7I_{23},\\ \nonumber
&&&&&&\\ \nonumber
K^{10}_{30} &= &I_{30}/3,&&
K^{12}_{30} &= &-\frac{5}{6}I_{30}, \\ \nonumber
K^{11}_{31} &= &3(K_{31} - \frac{1}{2}I_{31}),&&
K^{10}_{32} &= &\frac{5}{3}I_{32},\\ \nonumber
K^{12}_{32} &= &-\frac{25}{6}I_{32},&&
K^{11}_{33} &= &7(K_{33} - \frac{1}{2}I_{33}),\\ \nonumber 
&&&&&&\\ \nonumber
K^{20}_{20} &= &-I_{20}/3,&&
K^{22}_{20} &= &K_{20} - I_{20}/6, \\ \nonumber
K^{21}_{21} &= & -\frac{3}{2}I_{21},&&
K^{20}_{22} &= &-\frac{5}{3}I_{22}, \\ 
K^{22}_{22} &= &5(K_{22} - \frac{1}{6}I_{22}),&&
K^{21}_{23} &= &-\frac{7}{2}I_{23}.
\end{array}
\label{KIlIplp}
\end{equation}
where
\begin{eqnarray}
\nonumber
I_{20}& = &L_{20} - M_2 + N_2,\\ \nonumber
I_{21}& = &L_{21} + M_2 - N_2,\\ \nonumber
I_{22}& = &L_{22} - M_2 + N_2,\\ 
I_{23}& = &L_{23} + M_2 - N_2.\\ \nonumber
I_{30}& = &L_{30} + M_3 + N_3,\\ \nonumber
I_{31}& = &L_{31} - M_3 - N_3,\\ \nonumber
I_{32}& = &L_{32} + M_3 + N_3,\\ \nonumber
I_{33}& = &L_{33} - M_3 - N_3.\nonumber
\end{eqnarray}
Full analytical forms of the kernels $K_{ll'}$ are
\begin{eqnarray}   
\nonumber            
K_{20}  &= & 0, \\ \nonumber
K_{22}  &= & \frac{(s-4m_{\pi}^2) s (7 s-15 s'+32m_{\pi}^2)}{40 \pi  (s-s') (s'-4m_{\pi}^2)^2s'},\\ 
K_{31}  &= & \frac{s}{8\pi s'(4m_{\pi}^2 - s')}, \\ \nonumber
K_{33}  &= & \frac{s}{56\pi s'(4m_{\pi}^2 - s')^3(s - s')} \\ \nonumber
 &\times &
[-512m_{\pi}^6 + 8s^3 - 64m_{\pi}^4(s - 7s') - 35s^2s' \\ \nonumber
&+& 42ss'^2 - 7s'^3 + m_{\pi}^2(44s^2 - 56ss' - 84s'^2)].
\end{eqnarray}
The other (even nonzero) $K_{ll'}$ elements are not given here as they are multiplied, in
(\ref{KTIl}),
by zero $f^{I}_l(s)$ amplitudes.
It is caused by the Bose symmetry for two pions according to which 
the sum $I+l$ must be even. 
It also leads to the same condition for sum $l+l'$ in the equation (\ref{eq:DROSDR}).
The diagonal kernels $K_{22}(s,s')$ and $K_{33}(s,s')$ contain singularity at $s = s'$
which is the only type of singularities in presented equations.

In order to simplify the analytical forms of the $I_{ll'}$ elements, some terms 
which appear at least twice or help to reduce these formulas 
are given below:
\begin{widetext}
\begin{eqnarray}
\nonumber
a_1 &=& s -4m_{\pi}^2,\sps 
a_1' = s' -4m_{\pi}^2,\sps 
a_2 = 2s'-8m_{\pi}^2, \sps
a_3 = s + 2s'-4m_{\pi}^2,\sps
a_4 = 2s' - s -4m_{\pi}^2, \\ \nonumber
L_1 &=& \log \left(\frac{s'}{s + s' -4m_{\pi}^2}\right),\sps 
L_2 = \log \left(\frac{2s'}{s + 2s' - 4m_{\pi}^2}\right),\sps  
L_3 = \log \left(\frac{s + 2s' - 4m_{\pi}^2}{2(s + s' -4m_{\pi}^2)}\right), \\  
L_4 &=& \log \left(\frac{2s'-8m_{\pi}^2}{- s + 2s'-4m_{\pi}^2}\right), \sps 
L_5 = \log \left(2(-4m_{\pi}^2 + s + s')\right),
\end{eqnarray}

\begin{eqnarray}
\nonumber
f_1 &=& 16m_{\pi}^4 + s^2 + 8m_{\pi}^2(2s - 3s') - 6ss' + 6s'^2, \\ \nonumber
f_2 &=& 16m_{\pi}^4 + s^2 + 6ss' + 6s'^2 - 8m_{\pi}^2(s + 3s'), \\ \nonumber
f_3 &=& 64m_{\pi}^6 + s^3 + 48m_{\pi}^4(3s - 4s') - 12s^2s' 
+ 30ss'^2 - 20s'^3 + 12m_{\pi}^2(3s^2 - 12ss' + 10s'^2), \\
f_4 &=& 64m_{\pi}^6 - s^3 - 12s^2s' - 30ss'^2 - 20s'^3 
- 48m_{\pi}^4(s + 4s') + 12m_{\pi}^2(s^2 + 8ss' + 10s'^2), \\ \nonumber
f_5 &=& 192m_{\pi}^6 - 3s^3 - 148s^2s' + 600ss'^2 - 480s'^3 
- 16m_{\pi}^4(9s + 28s') + 4m_{\pi}^2(9s^2 - 304ss' + 360s'^2), \\ \nonumber
f_6 &=& 
-64 m_{\pi}^6+48 (s+4 {s'}) m_{\pi}^4+s^3+20 {s'}^3+30 s {s'}^2
-12 \left(s^2+8 {s'} s+10 {s'}^2\right) m_{\pi}^2+12 s^2 {s'}, \\ \nonumber
f_7 &=& \frac{{a_1}}{{a_1'}}\Big[9280 m_{\pi}^6-16 (287 s+748 \
{s'}) m_{\pi}^4+4 \left(139 s^2+896 {s'} s+1080 {s'}^2\right) m_{\pi}^2+3 \
s^3-480 {s'}^3-600 s {s'}^2-148 s^2 {s'}\Big]\\ \nonumber
&+&48 {f_4} L_2 + \frac{{a_1} {f_5}}{{s'}}. 
\end{eqnarray}

The analytical expressions of the  $I_{ll'}$ elements can then be expressed as
\begin{equation}
I_{20} = -\frac{2}{3 {a_1}^3 \pi }\big(-3 {a_1} {a_3}-{f_2} L_1
\big),
\end{equation}

\begin{eqnarray}
\nonumber
I_{21} &=& \frac{3}{{a_1}^3 {a_1'} \pi } \bigg\{2\, L_3 \bigg[64 m_{\pi}^6-16 (4 s+7 {s'}) m_{\pi}^4+4 \left(5 s^2+20 {s'} \
s+12 {s'}^2\right) m_{\pi}^2-2 s^3-6 {s'}^3-18 s {s'}^2-13 s^2 \
{s'}\bigg] \\
&+&
9 s {a_1}^2+24 {a_1'} m_{\pi}^2 {a_1}-12 s^2 {a_1}+72 m_{\pi}^2 s \
{a_1}-12 {a_1'} {s'} {a_1}-18 s {s'} {a_1} -2 {a_1'} {f_2} L_2\bigg\},
\end{eqnarray}

\begin{eqnarray}
\nonumber
I_{22} &=& -\frac{5}{6 \pi } \Bigg\{\frac{3 {a_1} \left(12 m_{\pi}^2-3 s-4 {s'}\right)+3 \
{a_1} \left(4 m_{\pi}^2+3 s-4 {s'}\right)+4 {f_1} L_4 -4 {f2} 
L_2
}{{a_1}^3} \\ 
&-&\frac{2 s }{{a_1'}^2}\Bigg[\frac{9 {s'}}{{a_1}}+
\frac{6 \left(32 m_{\pi}^4-8 (s+4 {s'}) m_{\pi}^2+s^2+7 \
{s'}^2+6 s {s'}\right)}{{a_1}^2}\\ \nonumber
&+&\frac{2}{{a_1}^3 s} \Bigg({a_1'}^2 \
{f_1} L_4 + L_3 \bigg(256 m_{\pi}^8-512 \
(s+{s'}) m_{\pi}^6+16 \left(19 s^2+52 {s'} s+19 {s'}^2\right) m_{\pi}^4 \\ \nonumber
&-&8 \left(9 s^3+43 {s'} s^2+43 {s'}^2 s+9 {s'}^3\right) m_{\pi}^2+6 \
s^4+6 {s'}^4+42 s {s'}^3+73 s^2 {s'}^2+42 s^3 \
{s'}\bigg) \Bigg)+3\Bigg]\Bigg\},
\end{eqnarray}

\begin{eqnarray}
\nonumber
I_{23} &=& \frac{7}{4 {a_1}^3 \pi } \Bigg\{6 {a_1} \left(4 m_{\pi}^2+3 s-4 {s'}\right) \\ \nonumber
&+&\frac{{a_1} s }{{a_1'}^3}\Big[5824 m_{\pi}^6-48 (79 s+168 {s'}) m_{\pi}^4+12 \left(95 s^2+348 {s'} \
s+240 {s'}^2\right) m_{\pi}^2-115 s^3-312 {s'}^3-792 s {s'}^2-660 \
s^2 {s'}\Big]\\ \nonumber
&+&\frac{8}{{a_1'}^3}L_3 \Big[1024 m_{\pi}^{10}-256 (14 s+9 \
{s'}) m_{\pi}^8+64 \left(55 s^2+108 {s'} s+27 {s'}^2\right) m_{\pi}^6\\ \nonumber
&-&16 \left(92 s^3+333 {s'} s^2+252 {s'}^2 s+37 {s'}^3\right) m_{\pi}^4+4 \
\left(70 s^4+384 {s'} s^3+531 {s'}^2 s^2+236 {s'}^3 s+24 \
{s'}^4\right) m_{\pi}^2\\ 
&-&20 s^5-6 {s'}^5-78 s {s'}^4-253 s^2 \
{s'}^3-312 s^3 {s'}^2-150 s^4 {s'}\Big] 
\\ \nonumber
&-&2 \left({a_1} \left(-36 \
m_{\pi}^2+9 s+12 {s'}\right)+4 {f_2} 
L_2
\right)\Bigg\},
\end{eqnarray}

\begin{eqnarray}
\nonumber
I_{30} &=& \frac{1}{72 \
{a_1}^4 {a_1'} \pi  {s'}} \Bigg\{
120 {a_1'} s {s'} {a_1}^2+\Bigg[{a_1'} \Bigg(192 \
m_{\pi}^6-16 (9 s+28 {s'}) m_{\pi}^4+4 \left(9 s^2+176 {s'} s+360 \
{s'}^2\right) m_{\pi}^2 -3 s^3\\ \nonumber
&-&480 {s'}^3-360 s {s'}^2-148 s^2 \
{s'}\Bigg)
+{s'} \Bigg(9280 m_{\pi}^6-16 (287 s+748 {s'}) m_{\pi}^4+4 \
\left(139 s^2+896 {s'} s+1080 {s'}^2\right) m_{\pi}^2\\ 
&+&3 s^3-480 \
{s'}^3-600 s {s'}^2-148 s^2 {s'}\Bigg)\Bigg]
{a_1} - f_6 48 {a_1'} {s'} L_1\Bigg\},
\end{eqnarray}

\begin{eqnarray}
\nonumber
I_{31} &=& \frac{1}{16 {a_1}^4 {a_1'} \pi  {s'}}
\Bigg\{
\log (s')\Big(-3072 {a_1'} {s'} m_{\pi}^6+9216 {a_1'} \
{s'}^2 m_{\pi}^4+2304 {a_1'} s {s'} m_{\pi}^4-5760 {a_1'} {s'}^3 \
m_{\pi}^2-4608 {a_1'} s {s'}^2 m_{\pi}^2 \\ \nonumber
&-&576 {a_1'} s^2 {s'} m_{\pi}^2+960 \
{a_1'} {s'}^4+1440 {a_1'} s {s'}^3+576 {a_1'} s^2 \
{s'}^2+48 {a_1'} s^3 {s'}\Big)
+48 {a_1'} {\log (2)} {s'} f_6\\\nonumber
&+&{a_1} \
\Bigg[{s'} \Bigg(-9280 m_{\pi}^6+16 (863 s+748 {s'}) m_{\pi}^4+4 \left(-331 \
s^2+240 {a_1} s-3296 {s'} s-1080 {s'}^2\right) m_{\pi}^2+93 s^3\\
&+&480 \
{s'}^3+2520 s {s'}^2+80 {a_1}^2 s+1108 s^2 {s'}-120 \
{a_1} s (s+3 {s'})\Bigg)\\  \nonumber
&+&{a_1'} \Bigg(-192 m_{\pi}^6+16 (9 s+28 \
{s'}) m_{\pi}^4-4 \left(9 s^2-304 {s'} s+360 {s'}^2\right) m_{\pi}^2+3 \
s^3+480 {s'}^3-600 s {s'}^2+148 s^2 {s'}\Bigg)\Bigg]\\ \nonumber
&-&48 \
{s'} \Big[256 m_{\pi}^8-64 (5 s+13 {s'}) m_{\pi}^6+48 \left(3 s^2+17 \
{s'} s+14 {s'}^2\right) m_{\pi}^4-4 \left(7 s^3+63 {s'} s^2+114 \
{s'}^2 s+50 {s'}^3\right) m_{\pi}^2\\ \nonumber
&+&2 s^4+20 {s'}^4+70 s \
{s'}^3+72 s^2 {s'}^2+25 s^3 {s'}\Big] L_5
-48 \left(4 m_{\pi}^2+{a_1'}-2 s-{s'}\right) {s'} f_6 \log ({a_3})
\Bigg\},
\end{eqnarray}

\begin{eqnarray}
I_{32} &=& 
\frac{5}{72 {a_1}^4 \pi } 
\Bigg\{f_7
-\frac{48 }{{a_1'}^2}\Big[1024 m_{\pi}^{10}-256 (9 s+14 \
{s'}) m_{\pi}^8+64 \left(27 s^2+108 {s'} s+55 {s'}^2\right) m_{\pi}^6\\ \nonumber
&-&16 \
\left(37 s^3+252 {s'} s^2+333 {s'}^2 s+92 {s'}^3\right) m_{\pi}^4+4 \
\left(24 s^4+236 {s'} s^3+531 {s'}^2 s^2+384 {s'}^3 s+70 \
{s'}^4\right) m_{\pi}^2\\ \nonumber
&-&6 s^5-20 {s'}^5-150 s {s'}^4-312 s^2 \
{s'}^3-253 s^3 {s'}^2-78 s^4 {s'}\Big] L_3
+\frac{6 {a_1} s }{{a_1'}^2}\\ \nonumber
&\times&\Big[5824 \
m_{\pi}^6-16 (73 s+668 {s'}) m_{\pi}^4+4 \left(93 s^2+736 {s'} s+1220 \
{s'}^2\right) m_{\pi}^2-31 s^3
-640 {s'}^3-820 s {s'}^2-388 s^2 \
{s'}\Big]
\Bigg\},
\end{eqnarray}
and
\begin{eqnarray}
I_{33} &=& 
-\frac{7}{48 \
{a_1}^4 \pi }
\Bigg\{f_7
-\frac{2 {a_1} s}{{a_1'}^3} \Bigg[96768 \
m_{\pi}^8-192 (277 s+1177 {s'}) m_{\pi}^6+16 \left(1513 s^2+8517 {s'} s+9348 \
{s'}^2\right) m_{\pi}^4\\ \nonumber
&-&4 \left(1145 s^3+10221 {s'} s^2+17076 \
{s'}^2 s+9540 {s'}^3\right) m_{\pi}^2+325 s^4+3360 {s'}^4+9300 s \
{s'}^3+10368 s^2 {s'}^2+4035 s^3 \
{s'}\Bigg]\\ \nonumber
&-&\frac{48 }{{a_1'}^3}L_3\Bigg[4096 m_{\pi}^{12}-15360 \
(s+{s'}) m_{\pi}^{10}+ 1445 s^3 {s'}^3+972 s^4 {s'}^2+270 s^5 {s'}+768 \left(23 s^2+67 {s'} s+23 {s'}^2\right) \
m_{\pi}^8\\ \nonumber
&-&192 \left(49 s^3+257 {s'} s^2+257 {s'}^2 s+49 \
{s'}^3\right) m_{\pi}^6+144 \left(18 s^4+141 {s'} s^3+253 {s'}^2 \
s^2+141 {s'}^3 s+18 {s'}^4\right) m_{\pi}^4\\ \nonumber
&-&36 \left(10 s^5+106 \
{s'} s^4+283 {s'}^2 s^3+283 {s'}^3 s^2+106 {s'}^4 s+10 \
{s'}^5\right) m_{\pi}^2+20 s^6+20 {s'}^6+270 s {s'}^5+972 s^2 \
{s'}^4\Bigg] 
\Bigg\}.
\end{eqnarray}
\end{widetext}


\end{document}